\documentclass[10pt, conference, compsocconf]{IEEEtran}
\ifCLASSINFOpdf
  \usepackage[pdftex]{graphicx}
\else
\fi
\usepackage{eso-pic}

\begin{document}

% \AddToShipoutPicture*{\small \sffamily\raisebox{1.8cm} {\hspace{1.8cm}978-1-4673-5828-6\/13\/\$31.00 \copyright2013 IEEE}}
\AddToShipoutPicture*{\small \sffamily\raisebox{1.8cm} {\hspace{1.8cm} 
Copyright \copyright~2013 IEEE}}

%
% paper title
% can use linebreaks \\ within to get better formatting as desired
\title{Performance Optimization of WSNs using External Information}

% author names and affiliations
% use a multiple column layout for up to two different
% affiliations

\author{\IEEEauthorblockN{Gabriel Martins Dias}
\IEEEauthorblockA{Department of Information and Communication Technologies, Universitat Pompeu Fabra, Barcelona, Spain.\\
Email: gabriel.martins@upf.edu}
}

% conference papers do not typically use \thanks and this command
% is locked out in conference mode. If really needed, such as for
% the acknowledgment of grants, issue a \IEEEoverridecommandlockouts
% after \documentclass

% for over three affiliations, or if they all won't fit within the width
% of the page, use this alternative format:
% 
%\author{\IEEEauthorblockN{Michael Shell\IEEEauthorrefmark{1},
%Homer Simpson\IEEEauthorrefmark{2},
%James Kirk\IEEEauthorrefmark{3}, 
%Montgomery Scott\IEEEauthorrefmark{3} and
%Eldon Tyrell\IEEEauthorrefmark{4}}
%\IEEEauthorblockA{\IEEEauthorrefmark{1}School of Electrical and Computer Engineering\\
%Georgia Institute of Technology,
%Atlanta, Georgia 30332--0250\\ Email: see http://www.michaelshell.org/contact.html}
%\IEEEauthorblockA{\IEEEauthorrefmark{2}Twentieth Century Fox, Springfield, USA\\
%Email: homer@thesimpsons.com}
%\IEEEauthorblockA{\IEEEauthorrefmark{3}Starfleet Academy, San Francisco, California 96678-2391\\
%Telephone: (800) 555--1212, Fax: (888) 555--1212}
%\IEEEauthorblockA{\IEEEauthorrefmark{4}Tyrell Inc., 123 Replicant Street, Los Angeles, California 90210--4321}}

% use for special paper notices
%\IEEEspecialpapernotice{(Invited Paper)}

% make the title area
\maketitle

\begin{abstract}
The goal of this work is to describe a self-management system that correlates data sensed by different Wireless Sensor Networks (WSNs) and adjusts the number of active nodes in each network to provide an appropriate amount of measurements. The architecture considers the factors that make the external data relevant to the local network, such as the distance between covered areas, the relation between the types of sensed data and the reliability of the measurements. As a result, the operation of each network will be tuned to trade-off the accuracy of the measurements and the power consumption.

\end{abstract}


\section{Introduction}
\label{sec:introduction}

Different types of WSNs may be deployed in similar areas in order to measure semantically linked data, for example, temperature and humidity. Therefore, these data could be combined to enhance the accuracy of different WSNs without increasing their measurement rate and keeping their levels of energy consumption low. 

According to \cite{so65223}, most WSN applications for fire detection use a fusion of sensed data, such as temperature and CO$_{2}$ levels. This is usually mapped using functions that provide a trustful correlation between different types of parameters. Tests done in \cite{Bahrepour2010} show that a WSN for fire detection that uses ionization and photoelectric sensors might increase its accuracy up to 45\% if the temperature was also considered in the detection algorithm. Additionally, the amount of data collected by a WSN can be tuned if the network has information from external sources about the probability that the event it is monitoring exists. In case of a low probability, the amount of data to collect can be reduced, hence saving the sensors' batteries.

In \cite{6084070}, some examples of collaborative WSNs are described. Those networks either use collaboration methods as (internal) solutions for some problems of WSNs, such as coverage, localization, energy conservation, and security, or they cooperatively monitor and control objects (i.e. providing information to other systems). This work is meant to provide the missing link between both definitions: how to reduce the problems of WSNs, such as coverage and energy conservation, by cooperatively monitoring the environment.

So far, no work has considered using data collected by external WSNs in order to improve the performance of internal WSNs. As a side note, there are some challenges when considering to achieve an effective collaboration between WSNs that were not built to work together, because they may not cover exactly the same area, the data sensed by their neighbors may not be tightly coupled to their own type of monitored data, or simply because they are under different ownership. To handle these problems, it is necessary to weight the relevance of the data received from external WSNs as well as consider possible malicious behavior.

The main goal of this work is to maximize the lifetime of WSNs without compromising the quality of the measurements taken. To achieve this, the proposed system will be capable of self-configuring its networks at runtime according to the changes in the environment, detected by the internal or external WSNs.

% \begin{figure}[!t]
% \centering
% \includegraphics[width=2.5in]{map.pdf}
% \caption{Different WSNs covering similar areas}
% \label{fig_map}
% \end{figure} 

%\input{use-cases}
%\input{architecture}
\section{Architectural Components}
\label{sec:procedures}
%\begin{figure}
%\centering
%\includegraphics[scale=0.4]{figures/autonomic-architecture.png}
%\caption{Monitor, Analyze, Plan and Execute functions}
%\label{fig:autonomicarchitecture}
%\end{figure}

\begin{figure}[!t]
\centering
\includegraphics[scale=0.55]{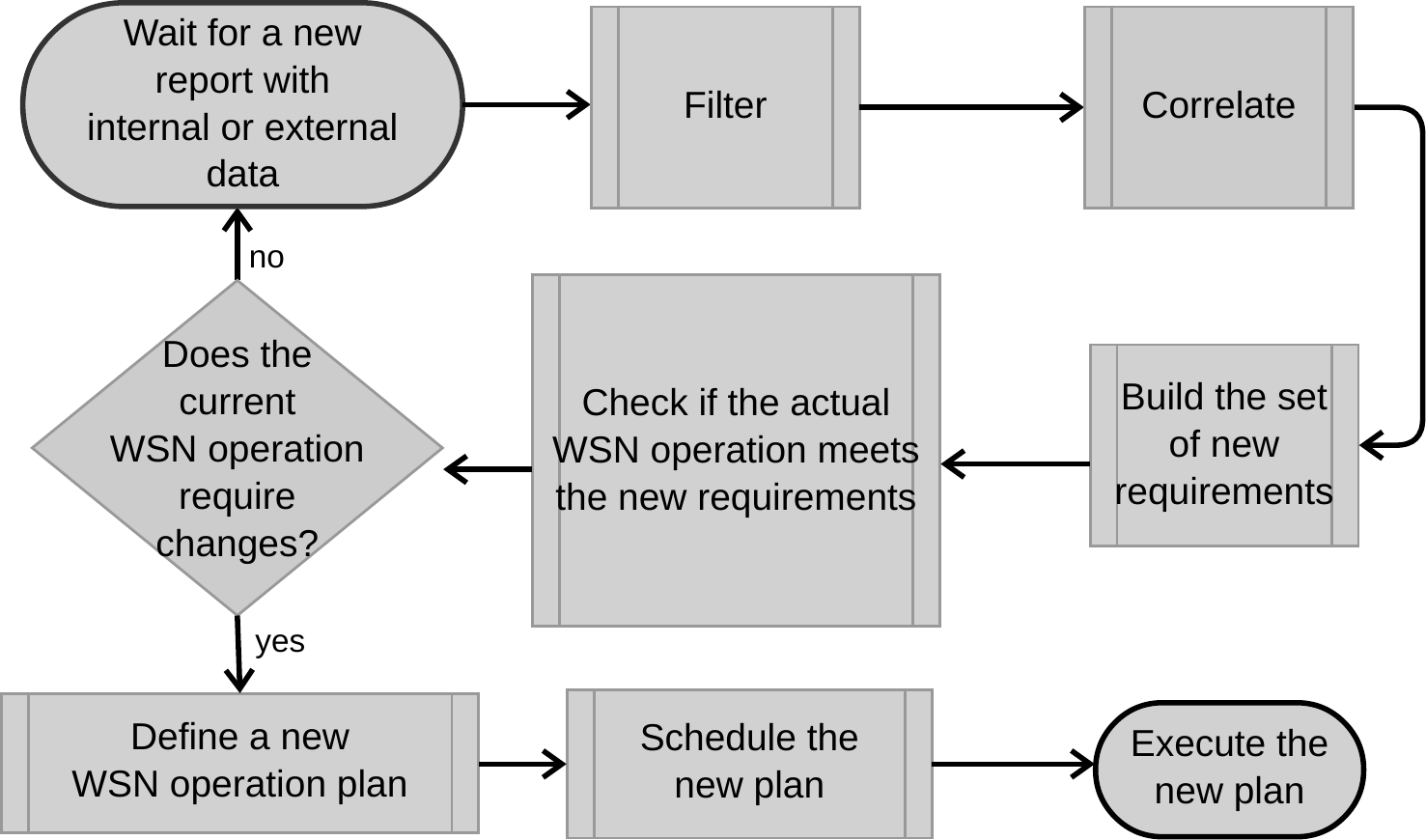}
\caption{Flowchart of an Enhanced Gateway (EG)}
\label{fig:flowchart}
\end{figure}

In \cite{Pal2012}, an architecture for WSN collaboration is presented. A new type of node, called an Enhanced Gateway (EG), is used as the contact point between the internal WSNs and the other EGs. They are directly connected to the respective WSN sink nodes and maintain an overlay connection to other EGs in order to exchange information about their networks and measured data. The same role is played by a manager in the autonomic systems (described in~\cite{citeulike:4167901}), which must perform monitor, analyze, plan and execute functions in order to configure its networks according to the environmental changes. Figure~\ref{fig:flowchart} shows the basic workflow. 

\subsection{Monitor}
After receiving new data either from internal or external WSNs, the EG must filter it in order to remove noise and errors by performing syntactic and semantic validations over the measurements. At this point, the WSN application goals (e.g. fire detection) must be considered in order to scale how relevant the data can be that is received from other WSNs. The filtered data will be used to find a relationship between different sensed data types.

\subsection{Analyze}
Every change in the environment must be analyzed in order to know what the probability is of having an event of interest, for example, a forest fire. The inference can be made either using pattern-matching or threshold-based algorithms.
 
Approaches that use manually-set thresholds require prior knowledge of the domain and the relationship between the different types of sensed data. In addition, each EG must keep information about the trust of the networks that are collaborating with it. The trust value may consider discrepancy to its own data, the distance between the networks and also the errors observed in past measurements. Artificial Intelligence techniques can be used to build a learning system that adjusts the trust value after evaluating each new measurement input. On the other hand, pattern-matching approaches require less external configuration and can build predictions based on historical data.
 
\subsection{Plan \& Schedule}

By checking the probability of having an event of interest, the EG will determine whether changes must be made in the network behavior. The new plan will depend on the type of the application deployed, which can be event-driven, monitoring or hybrid. Event-driven applications (e.g. decentralized fire detection) require no delay, quick transmissions and reliability. Monitoring applications (e.g. temperature and humidity) do continuous transmissions to the sink and tolerate packet losses as well as small delays between consecutive transmissions. Finally, hybrid applications (e.g. centralized fire detection) combine both types.

For example, if the network was operating with a low number of nodes transmitting measurement reports at a low rate, it might not manage to make an early detection of a new critical event such as a fire. According to the evaluation done, a request for changing the current set of active nodes and the time between consecutive reports will be issued.
 
In the end, the EG will compute the best combination of active nodes and the time interval between consecutive reports in case of continuous transmissions. The new plan must be a trade-off between the power consumption and the quality of measurements, which scaling may consider multiple parameters, such as the number of active nodes, the time interval between consecutive transmissions, network coverage, latency, packet loss rate, bandwidth used to report an event and information throughput at the sink.

\section{Performance Evaluation Plan}
\label{sec:evaluation}

Our first planned step is to simulate multiple WSNs that share and use their measured data to optimize their own operation. For example, temperature monitoring and fire detection. For that, the components described in Section~\ref{sec:procedures} will be implemented in the simulator, starting with a data correlation algorithm and the definition of how to quantify the quality of the measurements done in the WSNs. After this step, it will be possible to compare the energy consumption of the WSNs when collaboration is used versus the opposite case. A second step is to evaluate the proposed architecture in a test-bed, which will proof the suitability of the proposed approach in real world scenarios.

%Our first planned step is to simulate networks that may work with correlated data, for example, temperature monitoring and fire detection. Then, the modules of the managing system will be developed starting with a data correlation tool and the QoS parametrization followed by its evaluation. After these steps, it will be possible to produce the first results and compare them to the scenarios without collaboration. Additional functions to filter the measured data and aggregate its meaningful part may increase the quality of the system.

%The final system can use the same core implementation in the EGs added by an overlay communication between themselves, as well as the connection to the data storage, which will be modeled later. Then, it will be able to manage real WSNs and finally be evaluated by comparing the new energy consumption levels to those observed before its adoption.

%\section{Procedures}
\section{Expected Outcomes}
\label{sec:outcomes}

The main expected outcome of this work is an implementation of an autonomic system that provides a scalable and self-managed solution for WSNs management. 

The challenge is to build an architecture that will reduce the energy consumption of the WSNs based on the knowledge acquired from historical data. Additionally, the new method will use collaboration models and explore the WSNs as sources of knowledge to achieve this goal. 

Other expected outcomes are a generic approach for correlating data from different sources by applying pattern-matching or threshold-based algorithms, and metrics for measuring and evaluating the quality of measurements based on the application types.
%Also, the systems will be able to communicate and negotiate with systems from different owners, which may have WSNs deployed in the same monitored area.
%, increasing their lifetime , that may be either reports with complete information about the nodes and the measured data, or just a digest about past conditions.

\section{ACKNOWLEDGEMENT} 
I would like to acknowledge the contributions of my Ph.D. advisors, Simon Oechsner and Boris Bellalta, to this work, which has been partially supported by the Spanish Government, through the project TEC2012-32354.

\end{document}